\documentclass[twocolumn,pra,amsmath,amssymb,superscriptaddress]{revtex4-1}
\usepackage[utf8]{inputenc}
\usepackage{amsfonts,amsmath,amssymb}
\usepackage{graphicx}
\usepackage{bm} 
\usepackage{mathrsfs}
\usepackage{subfigure}
\usepackage{textcomp}
\usepackage{microtype}
\usepackage{hyperref}
\usepackage[flushleft]{threeparttable}
\usepackage[per-mode=symbol]{siunitx}
\usepackage[version=3]{mhchem}

\usepackage{booktabs}
\graphicspath{{../figures/}}

\DeclareSIUnit\intensity{\watt\per\centi\meter\squared}
\DeclareSIUnit\fieldstrength{\volt\per\centi\meter}
\usepackage{xspace}

\DeclareUnicodeCharacter{2009}{\,}

\newcommand{\degree}{\ensuremath{^\circ}}%

\newlength{\figwidth}
\newlength{\figwidthwide}
\let\orgautoref\autoref
\providecommand{\Autoref}{%
  \def\equationautorefname{Equation}%
  \def\figureautorefname{Figure}%
  \def\subfigureautorefname{Figure}%
  \def\tableautorefname{Table}
  \def\sectionautorefname{Section}%
  \orgautoref}
\renewcommand{\autoref}{%
  \def\equationautorefname{Eq.}%
  \def\figureautorefname{Fig.}%
  \def\subfigureautorefname{Fig.}%
  \def\sectionautorefname{Sec.}%
  \orgautoref}

\usepackage{color}
\usepackage[normalem]{ulem}
\definecolor{darkgreen}{rgb}{0.0,0.7,0.0}

\begin{document}

\title{Laser-induced Coulomb Explosion Imaging of Alkali Dimers on Helium Nanodroplets} 

\author{Henrik H. Kristensen}
\affiliation{Department of Physics and Astronomy, Aarhus University, Ny Munkegade 120, DK-8000 Aarhus C, Denmark}
\author{Lorenz Kranabetter}
\affiliation{Department of Chemistry, Aarhus University, Langelandsgade 140, DK-8000 Aarhus C, Denmark}
\author{Constant A. Schouder}
\affiliation{Department of Chemistry, Aarhus University, Langelandsgade 140, DK-8000 Aarhus C, Denmark}
\affiliation{Universit\'{e} Paris‐Saclay, CEA, CNRS, LIDYL, 91191 Gif‐sur‐Yvette, France}

\author{Jacqueline Arlt}
\affiliation{Department of Chemistry, Aarhus University, Langelandsgade 140, DK-8000 Aarhus C, Denmark}

\author{Frank Jensen}
\affiliation{Department of Chemistry, Aarhus University, Langelandsgade 140, DK-8000 Aarhus C, Denmark}

\author{Henrik Stapelfeldt}
\email[]{henriks@chem.au.dk}

\affiliation{Department of Chemistry, Aarhus University, Langelandsgade 140, DK-8000 Aarhus C, Denmark}

\date{\today}

\begin{abstract}
Alkali dimers, \ce{Ak2}, residing on the surface of He nanodroplets are doubly ionized due to multiphoton absorption from an intense, 50-fs laser pulse leading to fragmentation into a pair of alkali cations. Based on the measured kinetic energy distributions, $P(E_{\text{kin}})$, of the \ce{Ak^+} fragment ions, we retrieve the distribution of internuclear distances, $P(R)$, via the \ce{Ak2^{2+}} potential curve. Results are obtained for \ce{Na2}, \ce{K2}, \ce{Rb2}, and \ce{Cs2} in both the 1~$^1\Sigma_{g}^+$ ground state and in the lowest-lying triplet state 1~$^3\Sigma_{u}^+$, and for \ce{Li2} in the 1~$^3\Sigma_{u}^+$ state. For \ce{Li2}, \ce{K2}, and \ce{Rb2}, the center of the measured $P(R)$'s is close to the center of the wave function, $\Psi(R)$, of the vibrational ground state in the 1~$^1\Sigma_{g}^+$ and 1~$^3\Sigma_{u}^+$ states, whereas for \ce{Na2} and \ce{Cs2} small shifts are observed. For all the \ce{Ak2}, the width of the measured $P(R)$ is broader than $|\Psi(R)|^2$ by a factor of 2--4. We discuss that resonance effects in the multiphoton ionization and interaction of the \ce{Ak+} ion with the He droplet give rise to the observed deviations of $P(R)$ from $|\Psi(R)|^2$. Despite these deviations, we deem that timed Coulomb explosion will allow imaging of vibrational wave packets in alkali dimers on He droplets surfaces.

\end{abstract}

\maketitle

\section{Introduction}\label{sec:intro}

When a molecule in the gas phase is irradiated by a sufficiently intense fs laser pulse it undergoes multiple ionization. The resulting multiple charged molecular cation is typically unstable and breaks apart into ionic fragments, a process termed Coulomb explosion~\cite{yatsuhashi_multiple_2018}. The spatial orientation and structure of the molecule, at the instant the laser pulse arrives, are imprinted on the emission direction and on the kinetic energy of the ionic fragments. Therefore, measurements of these experimental observables provide an opportunity for extracting information about how the molecule is turned in space and about its structure at the instant the laser pulse ionizes the molecule~\cite{schouder-arpc,li_coulomb_2022}. In particular, for a two-body system such as a diatomic molecule, the distribution of internuclear distances, $P(R)$, can be determined from the measured distribution of kinetic energies, $P(E_{\text{kin}})$ of the atomic fragment ions provided there is a unique correspondence between $R$ and $E_{\text{kin}}$. This is fulfilled if the interaction between the fragment ions is characterized by a single potential curve like in the case of two \ce{H^+} ions (\ce{He^+} ions) produced from double ionization of a hydrogen molecule~\cite{legare_time-resolved_2003,ergler_spatiotemporal_2006} (He dimer~\cite{zeller_imaging_2016}) or for two atomic ions resulting from double ionization of a dissociating molecule, like \ce{I2}, where the internuclear distance is so large that there is no molecular bonding left~\cite{petersen_control_2004}. In many other cases, the multiply charged ions can, however, fragment via several different potential curves, which precludes an accurate retrieval of $P(R)$ in the parent diatomic molecule or molecular dimer. Recently, this was illustrated for double ionization of \ce{(CS2)2} and \ce{Ar2}~\cite{schouder_laser-induced_2020}.

Here we study femtosecond laser-induced Coulomb explosion of alkali homodimers, i.e. \ce{Ak2}, where \ce{Ak} is either \ce{Li}, \ce{Na}, \ce{K}, \ce{Rb}, or \ce{Cs}. Upon double ionization, \ce{Ak2} will be stripped of its two valence electrons and the resulting closed-shell structure of \ce{Ak2^{2+}} gives rise to only a single, repulsive potential curve. As such Coulomb explosion of alkali dimers appears well-suited for determining the distribution of their internuclear distances. The purpose of the current work is to explore if this is the case and how closely the determined $P(R)$ resembles the internuclear wave function. The alkali dimers are formed at the surface of nanometer sized droplets of liquid helium, which offers an opportunity for studying \ce{Ak2} in both the 1~$^1\Sigma_{g}^+$ ground state and in the lowest-lying triplet state 1~$^3\Sigma_{u}^+$ (often termed the $X$ state and the $a$ state, respectively)~\cite{stienkemeier_laser_1995, higgins_helium_1998, bruhl_triplet_2001, mudrich_formation_2004, lackner_spectroscopy_2013, kristensen_quantum-state-sensitive_2022}. Our motivation for the work is the fact that if Coulomb explosion is capable of retrieving $P(R)$, and thereby the internuclear wave function, it opens opportunities for directly imaging the time evolution of vibrational wave packets. In particular, given the presence of the He droplet, it should allow real-time observations of how coupling between the He environment and the dimer perturbs the coherence and perhaps the eigenstate populations of vibrational wave packets~\cite{schlesinger_dissipative_2010,gruner_vibrational_2011,thaler_long-lived_2020}.

\section{Principle of Coulomb explosion imaging of \ce{Ak2}}\label{sec:principle}

\begin{figure}
\includegraphics[width = 8.6cm]{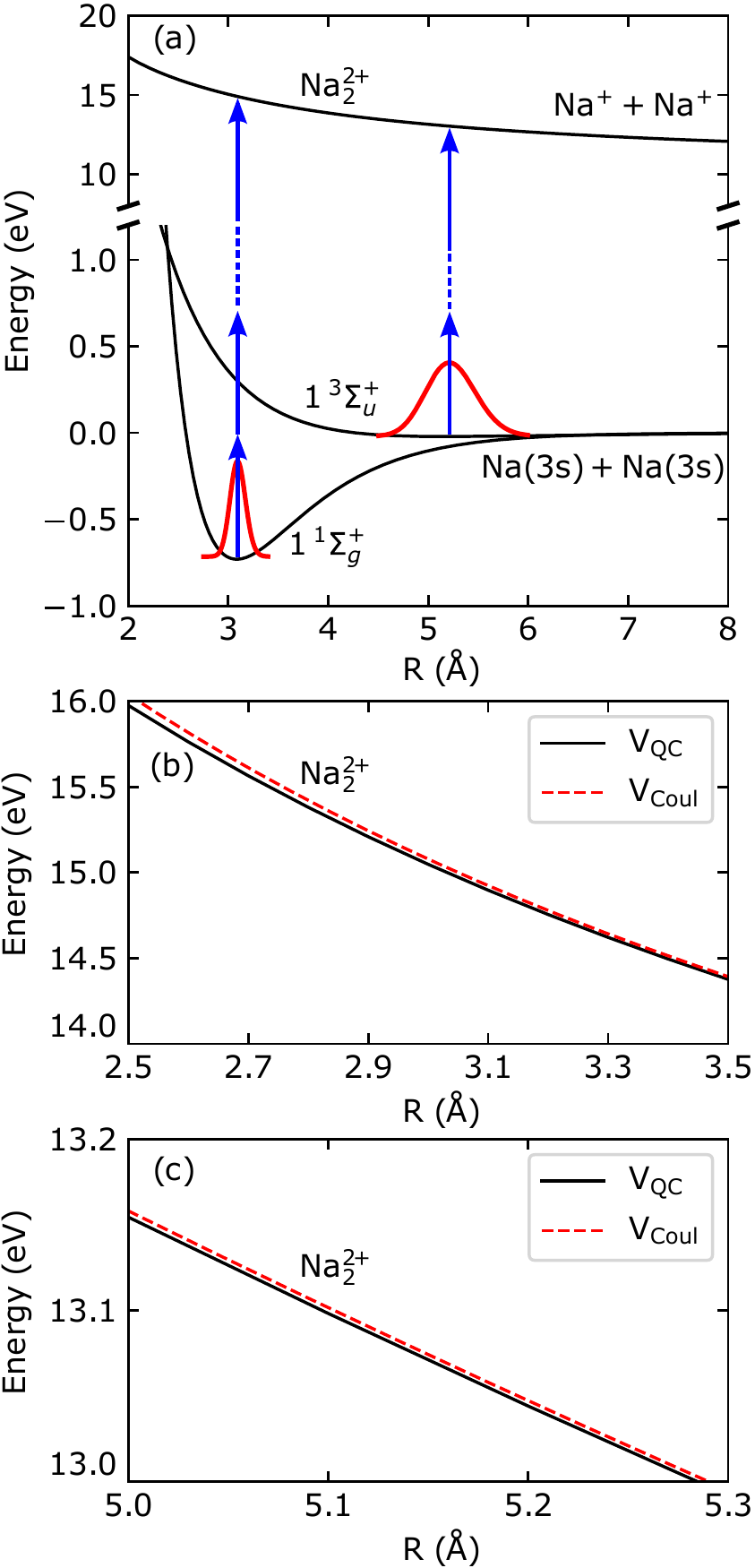}
\caption{(a) Energy diagram showing the potential curves for \ce{Na2} in the 1~$^1\Sigma_{g}^+$~\cite{magnier_potential_1993} and 1~$^3\Sigma_{u}^+$ states~\cite{bauer_accurate_2019} and $V_\text{QC}(R)$ for \ce{Na2^{2+}}. The square of the vibrational ground state wave functions for the 1~$^1\Sigma_{g}^+$ and 1~$^3\Sigma_{u}^+$ potentials are shown in red. The vertical blue arrows depict the laser photons (not to scale) to illustrate the double ionization process, which triggers Coulomb explosion of \ce{Na2}. (b)-(c) Zoomed-in view of the $V_\text{QC}(R)$ and $V_\text{Coul}(R)$ potentials for \ce{Na2^{2+}} around the equilibrium distance of the 1~$^1\Sigma_{g}^+$ and 1~$^3\Sigma_{u}^+$ states, respectively~\cite{deiglmayr_calculations_2008, bauer_accurate_2019}.}
\label{fig:Na2_potentials_sketch}
\end{figure}

The starting point of our method is double ionization of an alkali dimer \ce{Ak2} in either the 1~$^1\Sigma_{g}^+$ or in the 1~$^3\Sigma_{u}^+$ state. Assuming that the dimers have equilibrated to the 0.37 K temperature of the droplets~\cite{aubock_triplet_2007}, then only the vibrational ground state is populated in either of the two electronic states. The double ionization happens as a result of absorption of $N$ photons from an intense femtosecond laser pulse as illustrated in \autoref{fig:Na2_potentials_sketch}(a) for the case of \ce{Na2}, i.e.

\begin{align*}
	\ce{Na2} + N~h\nu &\rightarrow \ce{Na2^{2+}} + 2e^-.
\end{align*}
With a central wavelength of the laser pulse of 800 nm, absorption of at least 9 (10) photons is required to double ionize \ce{Na2} in the $^1\Sigma_{g}^+$ ($^3\Sigma_{u}^+$) state. The double ionization projects the vibrational wave function of \ce{Na2} onto the potential curve of \ce{Na2^{2+}}. The repulsive character of this curve causes the molecular dication to subsequently break apart into a pair of \ce{Na+} ions, i.e. Coulomb explosion due to the electrostatic repulsion between the two \ce{Na+} ions. During the Coulomb explosion, the potential energy of \ce{Na2^{2+}} is converted into kinetic energy of the two atomic fragments. Thus, a \ce{Na2^{2+}} ion with an internuclear separation of $R$ will produce two \ce{Na+} ions each with $E_{\text{kin}}$ = $\dfrac{1}{2}V(R)$, where $V(R)$ is the potential curve of \ce{Na2^{2+}}. In the experiment, we measure the distribution of kinetic energies, $P(E_{\text{kin}})$.  Due to the unique correspondence between $R$ and $V(R)$, we can determine the distribution of $R$, $P(R)$, by a standard probability distribution transformation with the appropriate Jacobian.  This procedure requires an explicit expression for $V(R)$. We use the result, $V_\text{QC}(R)$ illustrated in \autoref{fig:Na2_potentials_sketch}(a), from a quantum chemistry calculation, although the simple Coulomb potential $V_\text{Coul}(R)$ = 14.4 eV/$R$[\AA] provides a very good approximation in the pertinent $R$-interval around the equilibrium distance, $R_{\text{eq}}$, at 3.08 and 5.17 Å for the $^1\Sigma_{g}^+$ ($^3\Sigma_{u}^+$) state, respectively, see \autoref{fig:Na2_potentials_sketch}(b)-(c).

\section{Experimental setup}\label{sec:setup}

\subsection{Helium droplet machine}

\begin{figure}
\includegraphics[width=8.6 cm]{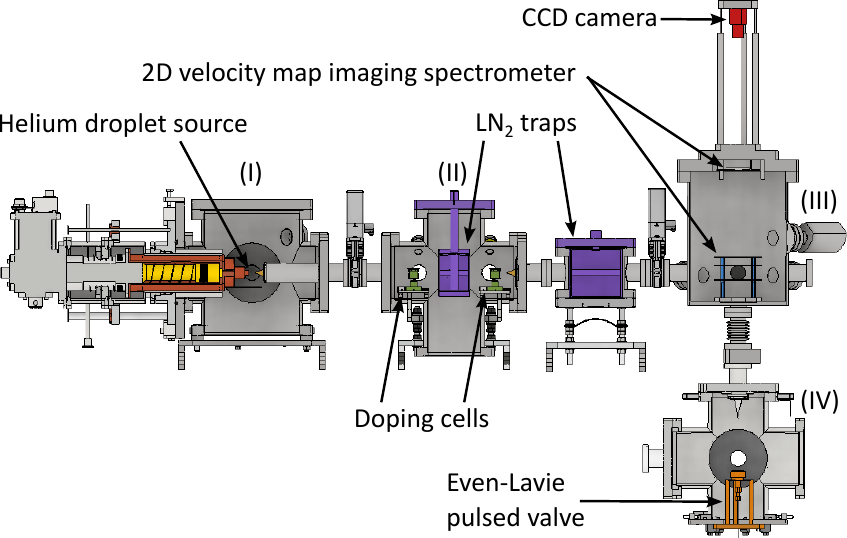}
\caption{Schematic drawing of the helium droplet machine. It consists of four vacuum chambers. The helium droplet source chamber (I), the doping chamber (II), the target chamber (III), and another source chamber (IV) equipped with a pulsed Even-Lavie valve. The target chamber contains a 2D velocity map imaging spectrometer.}
\label{fig:setup}
\end{figure}

\Autoref{fig:setup} shows a schematic drawing of the helium droplet machine used for the experiment. The machine consists of four vacuum chambers connected with pneumatic gate valves. The helium droplet experiments utilize three of these chambers: the helium droplet source chamber (I), the doping chamber (II), and the target chamber (III). The fourth chamber (IV) is a secondary source chamber containing a pulsed supersonic Even-Lavie valve (EL-7-4-2015-HRR, HT, 1 kHz).

A continuous helium droplet beam is created in the main source chamber (I). To achieve this, high purity helium gas is first lead through 2 meters of thin (1/16 in.) copper pipe wrapped around the first and the second stage of the coldhead of a closed cycle cryostat (Sumitomo Heavy Industries, RDK-415D), precooling the helium to around 4 K. Next, the helium gas is sent through a small nozzle assembly mounted on the tip of the coldhead. The nozzle assembly can be counter heated via a pair of heating resistors (Farnell, MHP35 470F) for effective temperature control. These are operated via a temperature controller (Lake Shore 335), which also monitors the temperature via a silicon diode (Lake Shore, DT-670B-CU). The nozzle assembly is clad with copper to shield it from heat radiation to ensure stable operation. In the present work, the nozzle is held at a fixed temperature $T_\text{nozzle}$ between 11 and 16 K. Ultimately the helium gas is subcritically expanded through a 5-$\mu\text{m}$-diameter nozzle (Platinum, 2 mm x 0.6 mm, PLANO) at a stagnation pressure of 25 bar into the source chamber. This expansion produces He droplets with a mean number of He atoms between $\sim$5000 and $\sim$14000 -- determined by $T_\text{nozzle}$~\cite{toennies_superfluid_2004}.

The helium droplet beam formed passes through a 1 mm skimmer (Beam Dynamics, Model 2) into the doping chamber (II). The doping chamber can be equipped with up to two different pickup cells to allow sequential doping of the He droplets with atoms or molecules. Each pickup cell consists of a crucible capable of holding a solid sample, a lid for the crucible, and a socket the crucible can be placed in. The crucibles and sockets have 3 mm holes in the sides to allow the He droplet beam to pass through. During the passage the He droplets can pick up gas-phase atoms, clusters or molecules that they collide with~\cite{toennies_superfluid_2004,barranco_helium_2006}. Kapton wires wrapped around the sockets can be resistively heated by sending current through in order to control the vapor pressure and thereby the doping probability in the pickup cells. PID controllers are used to monitor and control the heating via a set of thermocouple wires. To enable flexible and consistent positioning of the pickup cells the sockets are mounted on rails in the chamber. In this work, only the first pickup cell is installed and loaded with a solid sample of either \ce{Li}, \ce{Na}, \ce{K}, \ce{Rb}, or \ce{Cs}. The vapor pressure is regulated to a value where some of the droplets become doped with two \ce{Ak} atoms, which can then form a dimer on the surface~\cite{stienkemeier_use_1995, bruhl_triplet_2001}. Alternatively, pickup cells connected to an external gas reservoir can be used. In this way, gas phase atoms or molecules from the external reservoir are added through an ultra-fine needle valve (Kurt Lesker, VZLVM267) to the region where they can be picked up by the He droplets~\cite{shepperson_strongly_2017}.

After passing through the pickup cell, the now doped helium droplet beam passes through two coldtraps separated by a 2 mm skimmer (Beam Dynamics, Model 2). The coldtraps are filled with liquid \ce{N2} to reduce the effusive signal from undoped atoms that escape from the pickup cell. The beam is once more skimmed by a 2 mm skimmer (Beam Dynamics, Model 2), before reaching the detection chamber (III). Here the doped droplet beam is intersected by a pulsed laser beam in the center of a 2D velocity map imaging (VMI) spectrometer.

The secondary source chamber (IV) is placed beneath the detection chamber. This chamber is equipped with an Even-Lavie valve that can be used to create a pulsed molecular beam for gas-phase experiments via a supersonic expansion~\cite{arlt_photoelectron_2021}. The beam is directed through a skimmer (Beam Dynamics, Model 50.8) into the detection chamber, where it enters the VMI spectrometer from below. In this work, a molecular beam of gas-phase \ce{I2} was used for the energy calibration of the 2D imaging spectrometer.

The source chamber (I) is pumped by a 1900 l/s turbo molecular pump (Pfeiffer HiPace 2300), assisted by a 400 m$^3$/h backing pump setup (Edward Booster EH500 Hyd supported by an Edward E2M80 pump). The doping chamber is pumped via a 520 l/s (Pfeiffer TMU 521) and a 400 l/s (Leybold Turbovac 361) turbo molecular pump, both supported by a 16 m$^3$/h backing pump (Leybold Trivac D16B). The detection chamber (III) and the supersonic chamber (IV) are each pumped by a 685 l/s (Pfeiffer HiPace 700) turbo molecular pump. A 70 m$^3$/h backing pump (Pfeiffer Duo 65) assists the supersonic chamber pump, while the detection chamber pump is assisted by the same backing pump (Leybold Trivac D16B) supporting the doping chamber pumps.

\subsection{VMI spectrometer and laser}

A standard 2D VMI spectrometer~\cite{chandler_twodimensional_1987, parker_photoelectron_1997}, located in the detection chamber (III), is used to measure the velocity of ionic fragments created in our experiments. The VMI setup consists of an open three electrode electrostatic lens in combination with two stacked microchannel plates (MCPs - El Mul Technologies B050V, $\sim$40 mm active diameter) and a phosphor screen (El Mul Technologies Scintimax P47). The electrostatic lens is located inside a cylinder of mu-metal. The MCP and the phosphor screen are mounted on a flight tube located on top of the detection chamber. The flight tube improves the velocity resolution of low kinetic energy ions and helps separating ions with close mass-to-charge ratios. The chamber is equipped with multiple high voltage feedthroughs for the three electrodes, respectively labeled repeller, extractor and ground, the MCP (both for the front and the backside) and the phosphor screen. The phosphor screen is imaged with a CCD camera (Allied Vision Prosilica GE680) recording frames at 100 Hz, i.e. each frame contains data from 10 laser shots. A high voltage switch is used to gate the MCP such that only ions with a single mass-to-charge ratio are recorded at any given time.

The doped helium droplet beam is intersected perpendicularly by a pulsed, linearly polarized laser beam (from a Solstice Ace, Spectra-Physics, 1 kHz laser system) in the interaction region between the repeller and the extractor electrodes. The laser beam is focused by a 30 cm plano convex lens located in front of the detection chamber, leading to a focal spot size of about $\omega_0$ = 85~$\mu\text{m}$ (130~$\mu\text{m}$ for the \ce{Cs2} measurements). The duration of the pulses is $\sim$50 fs and their central wavelengths and intensities are given in~\autoref{fig:ion_energies}(a2)--(e2).

\section{Calculation of \ce{Ak2^2+} potential curves}

The $V_\text{QC}(R)$ potential energy curves for the \ce{Ak_2^2+} species were determined at the CCSD(T) level with a sequence of basis sets with increasing cardinal number in order to judge the basis set convergence. A counter-poise correction~\cite{van_duijneveldt_state_1994, boys_calculation_1970} was calculated for each internuclear distance, and the raw and counter-poise corrected energies for the two basis set with largest cardinal number were extrapolated to the basis set limit by an L$^{-3}$ formula~\cite{helgaker_convergence_1997}. The final $V_\text{QC}(R)$ potential energy curves were obtained by averaging the raw and counter-poise corrected extrapolated results~\cite{burns_counterpoise_2014, brauer_counterpoise_2014}.

The employed basis sets were: aug-cc-pCVXZ (X=D,T,Q) for \ce{Li} with all electrons correlated~\cite{prascher_gaussian_2011}, aug-cc-pCVXZ (X=D,T,Q,5) for \ce{Na} with the 1s-orbital treated as frozen core~\cite{prascher_gaussian_2011}, ANO-RCC-VXZP (X=D,T,Q) for \ce{K} with 5 orbitals frozen~\cite{roos_relativistic_2004}, ANO-RCC-VXZP (X=D,T,Q) for \ce{Rb} with 9 orbitals frozen~\cite{roos_relativistic_2004}, Sapporo AXZP (X=D,T,Q) in their DKH3 contracted version~\cite{noro_sapporo_2013} for \ce{Cs} with 18 core orbitals frozen and relativistic effects included by the Douglas-Kroll-Hess procedure to second order~\cite{douglas_quantum_1974, wolf_generalized_2002}. All calculations have been done using the Gaussian-16 program package~\cite{g16}.

The $V_\text{QC}(R)$ potential energy curves differ from the pure Coulombic by including exchange-repulsion and dispersion effects between the \ce{Ak+} ions. The $V_\text{QC}(R)$ are consequently slightly more repulsive at short distances and slightly less repulsive at long distances, compared to the pure Coulombic expression.

\section{Results}\label{sec:results}

\subsection{Ion images and kinetic energy distributions}
\label{sec:ion-images}

\begin{figure*}
\includegraphics[width=17.8 cm]{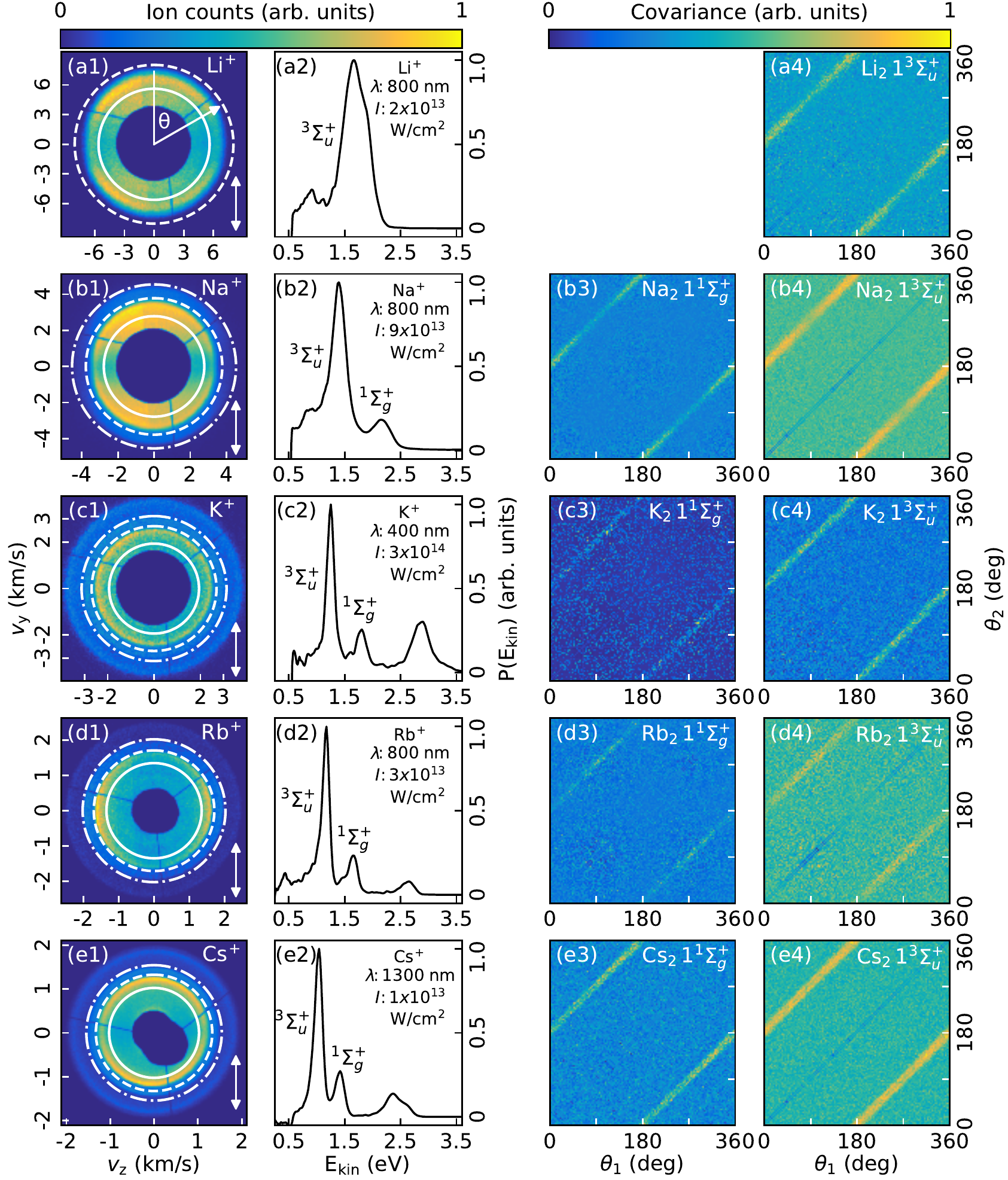}
\caption{(a1)--(e1) 2D velocity images of \ce{Ak+} ions. The white arrows in the bottom right corners indicate the polarization axis of the laser pulses. The annotated white rings designate the regions of the 1~$^3\Sigma_{u}^+$ and 1~$^1\Sigma_{g}^+$ channels, see text. (a2)--(e2) Kinetic energy spectra of the respective \ce{Ak+} ions. (b3)--(e3) [(a4)--(e4)] Angular covariance maps for the 1~$^1\Sigma_{g}^+$ [1~$^3\Sigma_{u}^+$] channel. $T_\text{nozzle}$ was 11 K for the data shown in (a)--(c), and 12 K for the data in (d) and (e).}
\label{fig:ion_energies}
\end{figure*}

\Autoref{fig:ion_energies}(a1)--(e1) show the 2D velocity images recorded for different alkali ions. Each image is obtained by stacking tens of thousands of frames. All images contain radial stripes and a center with no signal, see~\autoref{fig:ion_energies}(d1). The missing signal arises from a centrally positioned metal disk in front of the MCP and small mounting rods fixing it in place~\cite{schouder_laser-induced_2020, chatterley_laser-induced_2020}. The metal disk is used to block the many \ce{Ak+} ions, stemming from ionization of \ce{Ak} atoms in the effusive beam or on helium droplets, from reaching the detector. The center of the images in~\autoref{fig:ion_energies}(a1)--(c1) and (e1) have further been cut digitally to remove the few unwanted \ce{Ak+} ions passing by the edge of the central metal disk. This was done to improve the visual contrast of the outer features of the images. 

The kinetic energy distributions of the ions can be determined from the 2D velocity images. To do so, we apply an Abel inversion to the ion images to extract the radial velocity distribution. We specifically employ the polar onion peeling (POP) algorithm for the Abel inversion~\cite{roberts_toward_2009}. This algorithm retrieves the radial velocity distribution starting from the outer edges of the image and then moving to its center by subtracting the contribution from the high radii onto the lower ones. Therefore, the missing signal in the center of the images does not affect the retrieved velocity distributions of the data at higher radii. We use the polarization axis of the laser pulses as the axis of symmetry for the POP algorithm (see the white arrows annotated in~\autoref{fig:ion_energies}(a1)--(e1)). Finally, the kinetic energy distributions $P(E_{\text{kin}})$ are retrieved from the radial velocity distributions by applying the found energy calibration and the appropriate Jacobian transformation~\cite{schouder_laser-induced_2020}. \Autoref{fig:ion_energies}(a2)--(e2) show $P(E_{\text{kin}})$ for the corresponding ion images.

As discussed in our previous letter~\cite{kristensen_quantum-state-sensitive_2022}, each of the $P(E_{\text{kin}})$ distributions contains two peaks that can be assigned as the \ce{Ak+} ions stemming from Coulomb explosion of \ce{Ak2} in the 1~$^1\Sigma_{g}^+$ state or the 1~$^3\Sigma_{u}^+$ state. In \autoref{fig:ion_energies}(a2)--(e2) these peaks are labelled accordingly. The assignment is based on the close match between the central position of the observed peaks and the kinetic energy \ce{Ak+} ions would obtain through Coulomb explosion starting from the equilibrium position of either the 1~$^1\Sigma_{g}^+$ state or the 1~$^3\Sigma_{u}^+$ state. Note that for \ce{Li2} only the 1~$^3\Sigma_{u}^+$ state is observed, which is consistent with results from spectroscopic studies~\cite{higgins_helium_1998,lackner_spectroscopy_2013}. To support the interpretation that the ions in the two peaks are indeed produced by Coulomb explosion, we determined the angular covariance map of the \ce{Ak+} ions within the corresponding radial ranges in the images, i.e. between the dashed and dot-dashed circles for the 1~$^1\Sigma_{g}^+$ state and between the fully drawn and the dashed circles for the 1~$^3\Sigma_{u}^+$ state. An angular covariance map makes it possible to identify possible correlations in the emission direction of the ions both when they originate from gas phase molecules~\cite{hansen_control_2012,slater_covariance_2014,frasinski_covariance_2016,vallance_covariance-map_2021} and from molecules inside He nanodroplets~\cite{shepperson_strongly_2017,pickering_alignment_2018,schouder_structure_2019}. In each covariance map, displayed in \autoref{fig:ion_energies}(b3)--(e3) and (a4)--(e4), two diagonal lines centered at $\theta_2 = \theta_1 \pm \SI{180}{\degree}$  stand out and show that the emission direction of an \ce{Ak+} ion is correlated with another \ce{Ak+} ion departing in the opposite direction. Such a correlation identifies the ions as originating from the Coulomb explosion channel into an (\ce{Ak+}, \ce{Ak+}) pair.

In the experiments with \ce{K2}, \ce{Rb2} and \ce{Cs2} another channel is visible outside the dot-dashed rings marking the edge of the 1~$^1\Sigma_{g}^+$ channel, see~\autoref{fig:ion_energies}(c1)-(e1). We have found that these ions emanate from Coulomb explosion of equilaterally shaped \ce{K}, \ce{Rb} and \ce{Cs} trimers into three \ce{Ak+} ions~\cite{kranabetter_2021}. These ions make up the high energy peaks centered at 2.86 eV for \ce{K+}, 2.62 eV for \ce{Rb+} and 2.40 eV for \ce{Cs+}, as obtained by Gaussian fits, see~\autoref{fig:ion_energies}(c2)--(e2). If the trimers are only doubly ionized, they may produce \ce{Ak+} ions through the following dissociative ionization channels: \ce{Ak3^{2+}}~$\rightarrow$~\ce{Ak2^+}~+~\ce{Ak^+} and \ce{Ak3^{2+}}~$\rightarrow$~\ce{Ak^+}~+~\ce{Ak^+}+~\ce{Ak}. Based on the equilibrium geometries of the trimers, $E_{\text{kin}}$ of the \ce{Ak^+} ions from these channels fall outside the ranges of the \ce{Ak^+} ions from Coulomb explosion of the singlet and triplet dimers. Thus, we expect no pollution of the dimer signals from ionization of trimers. Experimentally, we confirmed this by recording data at lower vapor pressures in the doping cell. Under these conditions the peaks originating from Coulomb explosion of the trimers are strongly reduced, while no changes in the shape of $P(E_{\text{kin}})$ for the 1~$^1\Sigma_{g}^+$ state or for the 1~$^3\Sigma_{u}^+$ state were observed.

\subsection{Angular distributions}
\label{sec:angular-distribution}

\begin{figure}
\includegraphics[width = 8.6 cm]{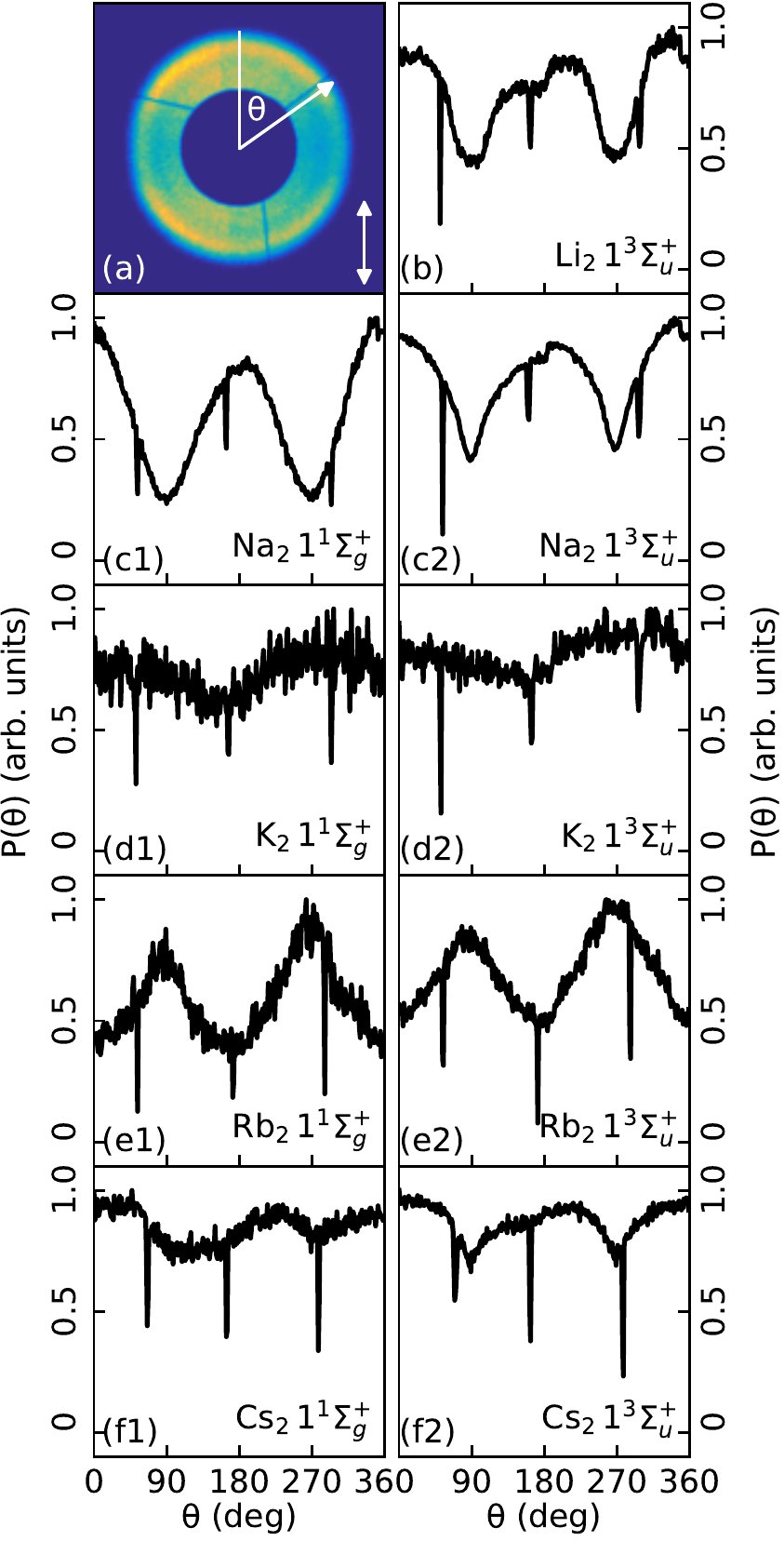}%
\caption{(a) Definition of the angle $\theta$ in the 2D velocity images. The white arrow in the lower right corner shows the laser polarization axis. (b) $P(\theta)$ distribution of the \ce{Li+} ions from the 1~$^3\Sigma_{u}^+$ region. (c1)--(f1), (c2)--(f2) $P(\theta)$ distributions of \ce{Na+}, \ce{K+}, \ce{Rb+}, and \ce{Cs+} ions from the 1~$^1\Sigma_{g}^+$ and 1~$^3\Sigma_{u}^+$ regions, see text.}
\label{fig:angular_distributions}
\end{figure}

The ion images in \autoref{fig:ion_energies} exhibit angular anisotropy. This is more clearly seen in the angular distributions, $P(\theta)$, obtained by integrating the ion images along the radial axis for the regions corresponding to the 1~$^1\Sigma_{g}^+$ and the 1~$^3\Sigma_{u}^+$ channel, respectively. \Autoref{fig:angular_distributions} shows that the emission directions of the \ce{Li+}, \ce{Na+} and \ce{Cs+} (\ce{K+} and \ce{Rb+}) ions are localized along (perpendicular to) the laser pulse polarization. We believe the angular anisotropy reflects the influence of electronically excited states in the multiphoton absorption process. In particular, the electronic structure of all alkali dimers is such that absorption of the first photon ($\lambda$ = 800 nm or 400 nm) is resonant or near-resonant with an electronically excited state. This can give rise to an alignment-dependent absorption probability depending on the transition dipole moment of the electronic states involved, and thereby an anisotropic angular distribution of the \ce{Ak+} fragments.

\begin{figure}
\includegraphics[width=8.6 cm]{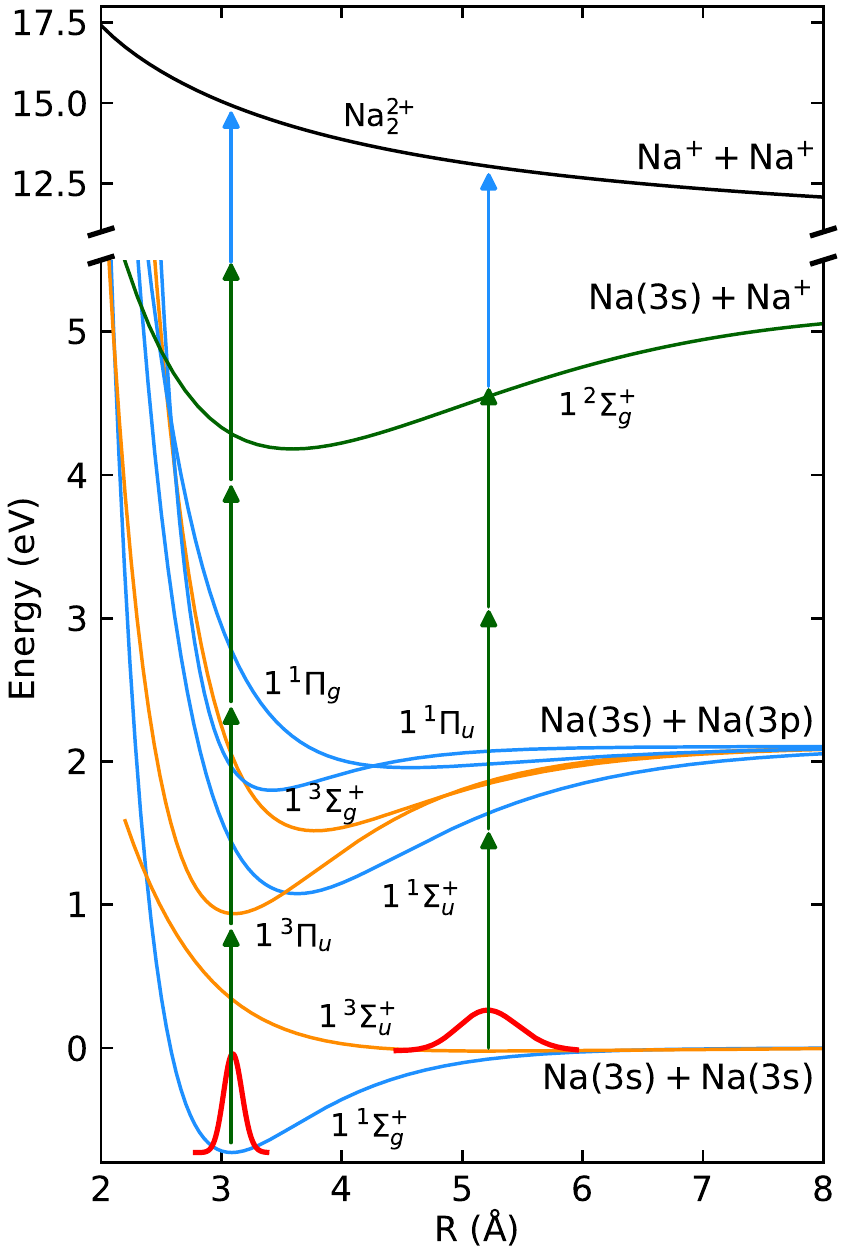}
\caption{Selected potential energy curves for the \ce{Na} dimer. Blue curves are singlet states, while orange curves are triplet states~\cite{magnier_potential_1993}. The green curve shows the 1~$^2\Sigma_{g}^+$ \ce{Na2^{+}} potential curve. The fully drawn black curve shows $V_\text{QC}(R)$ representing \ce{Na2^{2+}}. The red shapes show the square of the wave function of the vibrational ground state in the 1~$^1\Sigma_{g}^+$ and 1~$^3\Sigma_{u}^+$ potentials. The green arrows indicate laser photons (800 nm) in the multiphoton ionization process drawn to scale. The blue arrows illustrate the remaining laser photons involved in the double ionization process, not drawn to scale.}
\label{fig:Na2_curves}
\end{figure}

To illustrate the situation, we consider the case of \ce{Na2}. The potential energy diagram in \autoref{fig:Na2_curves} depicts a number of the lowest-lying electronic states. After absorption of the first photon ($\lambda$ =  800 nm), \ce{Na2} initially residing in the 1~$^1\Sigma_{g}^+$ state will be close to the 1~$^1\Sigma_{u}^+$ state. The 1~$^1\Sigma_{g}^+$~$\rightarrow$~1~$^1\Sigma_{u}^+$ transition is an allowed parallel transition meaning that the photon absorption probability is largest for the dimers with their internuclear axis parallel to the laser polarization axis. Upon further photon absorption, these parallel dimers will Coulomb explode and produce \ce{Ak^+} fragments emitted along the laser polarization because the \ce{Ak+} fragments recoil along the direction defined by the internuclear axis of their parent dimer. Thus, the parallel character of the transition will resonantly enhance the multiphoton process of \ce{Na2} (1~$^1\Sigma_{g}^+$) leading to double ionization and, thereby, it can explain why $P(\theta)$ for the 1~$^1\Sigma_{g}^+$ state is localized around $\SI{0}{\degree}$ and $\SI{180}{\degree}$. Turning to sodium dimers in the 1~$^3\Sigma_{u}^+$ state, \autoref{fig:Na2_curves} shows that absorption of the first photon brings them close to the 1~$^3\Sigma_{g}^+$ state. Since 1~$^3\Sigma_{u}^+~\rightarrow~1~^3\Sigma_{g}^+$ is also an allowed parallel transition, the same arguments apply to explain why multiphoton-induced double ionization of \ce{Na2} (1~$^3\Sigma_{u}^+$) produces \ce{Na^+} fragments emitted along the laser polarization. Besides the low-lying states discussed here, it is possible that higher-lying states in \ce{Na2} and/or states in \ce{Na2^+} can also influence the \ce{Na+} angular distributions.

Similar considerations were applied to the other alkali dimers and we found that the angular distibutions of the \ce{Ak+^} observed can also be accounted for by the parallel or perpendicular character of one-photon transitions from the 1~$^1\Sigma_{g}^+$ state or from the 1~$^3\Sigma_{u}^+$ state. Notably for \ce{K2}, where the wavelength of the laser pulse was 400 nm, the potential energy diagram~\cite{magnier_theoretical_2004} shows that absorption of the first photon can occur through both parallel and perpendicular transitions. Contribution to double ionization from these two transitions will lead to a reduced anisotropy in the angular distributions, a behavior consistent with the experimental measurements. No further details will be reported here. The bottom line is that the \ce{Ak^+} angular distributions strongly suggest that at the wavelengths of the laser pulses applied, resonance effects due to electronically excited states influence the multiphoton absorption process.

\subsection{Internuclear distributions}
\label{sec:internuclear-distribution}

\begin{figure}
\includegraphics{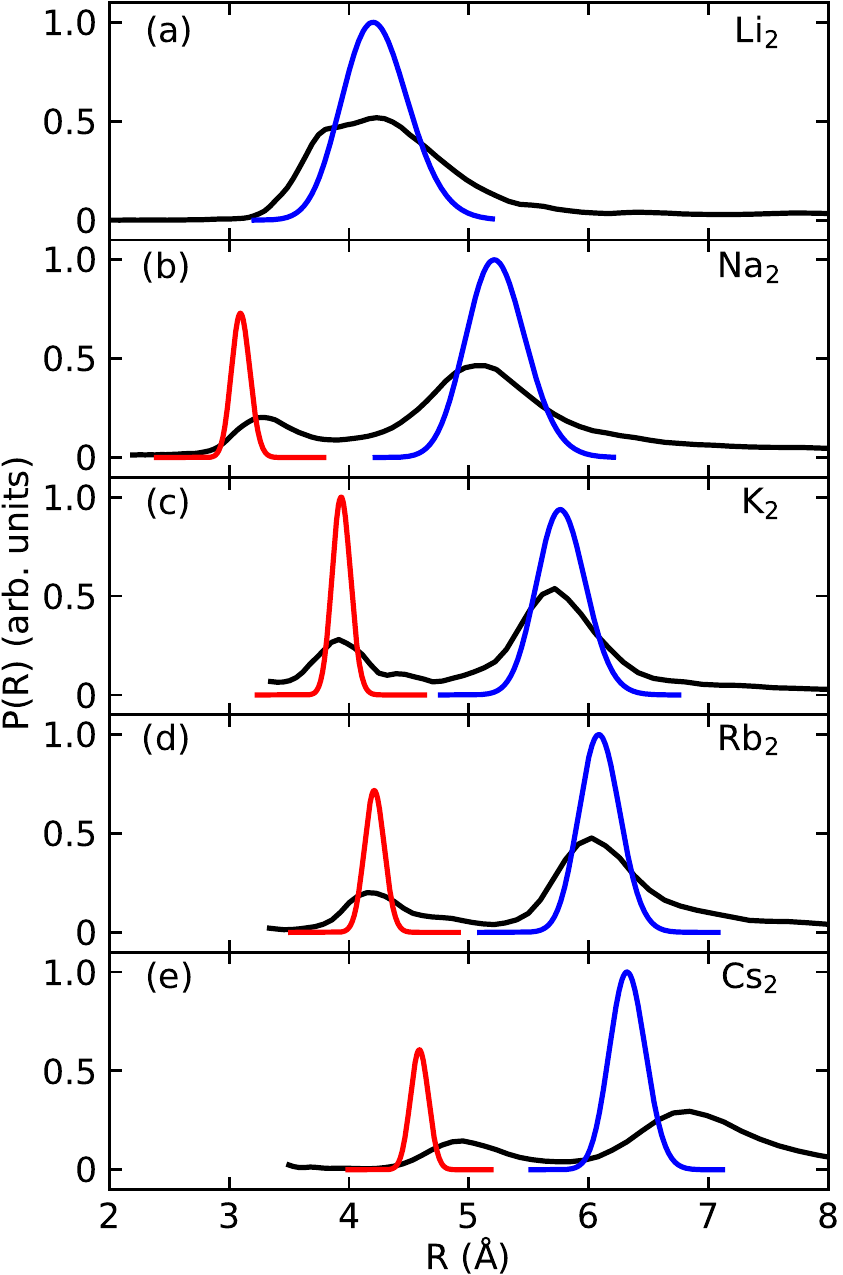}
\caption{$P(R)$ distributions for the \ce{Ak} dimers. The red (blue) shapes indicate the theoretical $|\Psi(R)|^2$ distributions for isolated 1~$^1\Sigma_{g}^+$ (1~$^3\Sigma_{u}^+$) states.}
\label{fig:wf_comparisons}
\end{figure}

As mentioned in \autoref{sec:principle}, the kinetic energy of an \ce{Ak^+} ion from Coulomb explosion of a bare \ce{Ak2} is given by $E_{\text{kin}}$ = $\dfrac{1}{2}V(R)$. Since $V(R)$ is a strictly decreasing function on the $R$-interval from 0 to $\infty$, we can determine the internuclear distribution, $P(R)$, by the following transformation of probability distributions: $P(R) = P(E_{\text{kin}})\left|\dfrac{dE_{\text{kin}}}{dR}\right| = P(E_{\text{kin}}) \left|\dfrac{1}{2}\dfrac{dV(R)}{dR}\right|$. Here $P(E_{\text{kin}})$ comes from the measurement, \autoref{fig:ion_energies}, and for $V(R)$ we use $V_\text{QC}(R)$, illustrated in \autoref{fig:Na2_potentials_sketch}(a). The black curves in \autoref{fig:wf_comparisons} show the resulting $P(R)$ for the five different alkali dimers. As references we display the calculated $|\Psi(R)|^2$ for the vibrational ground state in both the 1~$^1\Sigma_{g}^+$ and in the 1~$^3\Sigma_{u}^+$ state. The calculations were done by solving the vibrational stationary Schr\"{o}dinger equation for the isolated dimers using internuclear potentials from  the literature~\cite{magnier_potential_1993, magnier_theoretical_2004, jasik_calculation_2006, allouche_transition_2012, bauer_accurate_2019}. To ease the comparison between $P(R)$ and $|\Psi(R)|^2$, the latter were normalized to the experimental data. In practise, this was obtained by fitting $P(R)$ to the sum of two Gaussians, representing the probability distribution for the singlet and the triplet state, to ensure agreement in the overlapping region. The $|\Psi(R)|^2$ distributions were then individually normalized to the area of the corresponding $P(R)$ peak extracted from the fit. For \ce{Li2} only a single Gaussian was used for the fit.

\Autoref{fig:wf_comparisons} shows that $P(R)$ is broader than $|\Psi(R)|^2$ for all \ce{Ak2} in both the singlet and in the triplet state. From \autoref{tab:PR_fits}, summarizing the full width at half maximum (FWHM) of the experimental and calculated $R$-distributions, it can be seen that the FWHM of $P(R)$ for \ce{Li2}, \ce{Na2}, \ce{K2} and \ce{Rb2} in the 1~$^3\Sigma_{u}^+$ state is about a factor of two (1.8--2.4) times larger than the FWHM of the corresponding $|\Psi(R)|^2$. For \ce{Na2}, \ce{K2} and \ce{Rb2} in the 1~$^1\Sigma_{g}^+$ state the ratio between the FWHM of the experimental and calculated distributions is about four (3.9--4.4). For the 1~$^3\Sigma_{u}^+$ (1~$^1\Sigma_{g}^+$) state of \ce{Cs2} the factor is 3.7 (4.4).

\Autoref{tab:PR_fits} also lists the central position (CP) of $P(R)$ and of $|\Psi(R)|^2$. For \ce{Li2}(1~$^3\Sigma_{u}^+$) and \ce{K2} and \ce{Rb2} in either the 1~$^3\Sigma_{u}^+$ or in the 1~$^1\Sigma_{g}^+$ state, the CP of $P(R)$ deviates by $\leq$ 0.04~{\AA} from the CP of $|\Psi(R)|^2$. For \ce{Na2} the CP of $P(R)$ for the 1~$^1\Sigma_{g}^+$ (1~$^3\Sigma_{u}^+$) state is shifted by 0.22~{\AA} (-~0.09~{\AA}) compared to the CP of the corresponding $|\Psi(R)|^2$. For \ce{Cs2} the CP of $P(R)$ is displaced to higher values compared to the CP of $|\Psi(R)|^2$ for both the 1~$^1\Sigma_{g}^+$ state (0.38~{\AA)} and the 1~$^3\Sigma_{u}^+$ state (0.58~{\AA}).

\begin{ruledtabular}
\begin{table}
\centering
\caption{Central position and full width at half maximum (FWHM) for the peaks in the measured $P(R)$ and the calculated $|\Psi(R)|^2$ -- obtained via Gaussian fits. Up to about 15\% of the measured $P(R)$ FWHM can be explained by the experimental energy resolution of 100 meV (see text).}
\label{tab:PR_fits}
\begin{tabular}{lcccc}
      & \multicolumn{2}{l}{Central position (\AA)} & \multicolumn{2}{l}{FWHM  (\AA)} \\ \midrule
      & $P(R)$                  & $|\Psi(R)|^2$                 & $P(R)$         & $|\Psi(R)|^2$         \\ \midrule
\ce{Na_2} $^1\Sigma_g^+$ & 3.32   & 3.10     & 0.7          & 0.18                \\
\ce{K_2}  $^1\Sigma_g^+$ & 3.96   & 3.94     & 0.8          & 0.18                \\
\ce{Rb_2} $^1\Sigma_g^+$ & 4.25   & 4.21     & 0.8          & 0.19                \\
\ce{Cs_2} $^1\Sigma_g^+$ & 4.97   & 4.59     & 0.8          & 0.18                \\
\\
\ce{Li_2} $^3\Sigma_u^+$ & 4.22   & 4.22     & 1.2          & 0.65				  \\	
\ce{Na_2} $^3\Sigma_u^+$ & 5.13   & 5.22     & 1.4          & 0.58                \\	
\ce{K_2}  $^3\Sigma_u^+$ & 5.73   & 5.77     & 0.9          & 0.48                \\	
\ce{Rb_2} $^3\Sigma_u^+$ & 6.09   & 6.09     & 0.9          & 0.41                \\	
\ce{Cs_2} $^3\Sigma_u^+$ & 6.90   & 6.32     & 1.3          & 0.35                \\
\end{tabular}
\end{table}
\end{ruledtabular}

\section{Discussion}\label{sec:discussion}

We believe there are three factors in the experiment that can cause the experimentally retrieved $P(R)$ to deviate from the theoretical $|\Psi(R)|^2$. The first is the double ionization process. As explained in \autoref{sec:angular-distribution}, electronically excited states of \ce{Ak2}, and possibly also electronic states of \ce{Ak2^+}, can be transiently excited during the multiphoton absorption. The 50 fs duration of the laser pulse is long enough that the lighter dimers may undergo some internuclear motion on the resonant potential curves. Using again \ce{Na2} as an example, \autoref{fig:Na2_curves} shows that at the equilibrium distance of the initial 1~$^1\Sigma_{g}^+$ (1~$^3\Sigma_{u}^+$) state, the electronically excited singlet (triplet) states have a slope that will cause an increase (decrease) of $R$ upon internuclear motion. The same is true if \ce{Na2^+}(1~$^2\Sigma_{g}^+$) is transiently populated. This is consistent with the experimental observation that the CP of $P(R)$ is larger (smaller) for the 1~$^1\Sigma_{g}^+$ (1~$^3\Sigma_{u}^+$) state, see \autoref{fig:wf_comparisons}. Likewise, for \ce{Li2}, the shoulder of $P(R)$ towards smaller $R$-values than those spanned by $|\Psi(R)|^2$, is consistent with internuclear motion on potential curves of excited \ce{Li2} states or on the potential curve for \ce{Li2^+}(1~$^2\Sigma_{g}^+$). The larger masses of \ce{K2}, \ce{Rb2} and \ce{Cs2} are expected to reduce the internuclear motion on the intermediate potential curves, an expectation consistent with the very small deviation of the CP of the measured $P(R)$ from $|\Psi(R)|^2$ observed for \ce{K2} and \ce{Rb2}. For \ce{Cs2} the expectation does not hold, which we comment on below. We note that the internuclear motion may also contribute to a broadening of $P(R)$ with respect to $|\Psi(R)|^2$ and not just a shift. Finally, we point out that the passage through resonant states may also lead to a slight skew of the internuclear distribution due to Franck Condon overlaps and could be partly responsible for e.g. the small offset of $P(R)$  from $|\Psi(R)|^2$ for the 1~$^3\Sigma_{u}^+$ state of \ce{Na2}.

The second factor that can cause $P(R)$ to deviate from $|\Psi(R)|^2$ is the interaction between the recoiling \ce{Ak+} fragment ions and the He atoms on the droplet surface. Ongoing simulations based on time-dependent density functional theory show that the interaction induce a slight bending of the \ce{Ak+} ion trajectories and possibly a small loss of their kinetic energy. For a given dimer, like \ce{Rb2}, the magnitude of these effects are expected to be different for the singlet and the triplet state due to the different angular distributions of the dimer on the surface~\cite{guillon_theoretical_2011}. If the dimer lie at an angle to the surface, one \ce{Ak+} fragment might interact strongly with the helium, slowing it down. The partner fragment would then receive more than half the available Coulomb energy, leading to a broadening of $P(R)$. Furthermore, the effects are expected to be alkali-dependent. In particular, the observed shift of $P(R)$ for \ce{Cs2} may be a result of the \ce{Cs^+}-He interaction because the low recoil velocity of the \ce{Cs^+} ions, due to the large mass and small Coulomb energy, extends the interaction time compared to that of the fragments from the lighter \ce{Ak2}.

The third factor is the kinetic energy resolution of the VMI spectrometer. We estimate that the resolution is about 100 meV. This will broaden the kinetic energy spectra and thus lead to wider $P(R)$ distributions. Notably, such broadening will be largest for dimers with large internuclear distances, due to the shape of the \ce{Ak2^{2+}} potential curves. In particular for triplet state \ce{Rb2} and \ce{Cs2}, the broadening can account for up to about 15\% of the measured width of the $P(R)$ distributions.

\section{Conclusion and outlook}\label{sec:conclusion}

We used Coulomb explosion, induced through double ionization by an intense 50 fs laser pulse, of alkali homodimers on the surface of He nanodroplets, to determine the distribution of internuclear distances, $P(R)$. The agreement of the measured $P(R)$ with the theoretically expected internuclear distribution, i.e. $|\Psi(R)|^2$ for the vibrational ground state in either the 1~$^1\Sigma_{g}^+$ state or in the 1~$^3\Sigma_{u}^+$ state, is best for \ce{Li2}, \ce{K2}, and \ce{Rb2} in the triplet state. For these three dimers, the center of $P(R)$ lie within 0.04~{\AA} of the center of $|\Psi(R)|^2$ and the FWHM of $P(R)$ is about twice as large as the FWHM of $|\Psi(R)|^2$. As such, we deem our Coulomb explosion method capable of measuring time-dependent wave functions resulting from the creation of vibrational wave packets in \ce{Li2}(1~$^3\Sigma_{u}^+$), \ce{K2}(1~$^3\Sigma_{u}^+$), or \ce{Rb2}(1~$^3\Sigma_{u}^+$) with a precision of 0.04~{\AA} on the central position and a 1~{\AA} resolution of the shape.

One approach to creating vibrational wave packets is to use a moderately intense fs pump laser pulse to form a coherent superposition of vibrational eigenstates in the 1~$^3\Sigma_{u}^+$ state through the dynamic Stark effect~\cite{townsend_stark_2011,christensen_dynamic_2014,shu_femtochemistry_2017,claas_wave_2006}. An advantage of this nonlinear excitation scheme is its high excitation efficiency, which eliminates the need for subtraction of signal from unexcited dimers as is otherwise necessary when 1-photon excitation schemes are used~\cite{stapelfeldt_time-resolved_1998,yang_diffractive_2016}. The large magnitude and pronounced $R$-dependence of the polarizability of alkali dimers~\cite{deiglmayr_calculations_2008} make them particularly suited for dynamic Stark excitation. Results from ongoing experiments in our laboratory along these lines are very promising.

We believe Coulomb explosion imaging of alkali dimers on He droplets may be improved by the following measures. The first is to use shorter, mid-infrared laser pulses to induce ionization in the tunneling regime~\cite{wolter_strong-field_2015} and thus reduce the influence of resonances. The second is to implement a VMI spectrometer with a better energy resolution to eliminate any possible instrument broadening of the peaks in the measured kinetic energy distributions. Thereby, the resolution of the shape of the wave functions could be significantly enhanced. Finally, it will be necessary to develop a theoretical understanding of how the interaction between the He droplet and the alkali fragment ions influence the kinetic energy of the latter. Ongoing simulations are addressing this point~\cite{nadine_2022}.

\begin{acknowledgments}
We thank Jan Th{\o}gersen for expert help on keeping the laser system in optimal condition. H.S. acknowledges support from Villum Fonden through a Villum Investigator Grant No. 25886. The numerical  results presented in this work were obtained at the Centre for Scientific Computing, Aarhus https://phys.au.dk/forskning/faciliteter/cscaa/
\end{acknowledgments}

%



\end{document}